\documentclass[a4paper]{article}
\usepackage[latin1]{inputenc} %
\usepackage[T1]{fontenc} %
\usepackage{RR, amsmath, amssymb,amsbsy}
\usepackage{hyperref, subfigure,epsfig,graphics }

\RRdate{Janvier 2010}

\RRauthor{
Taous-Meriem~Laleg-Kirati \thanks{Taous-Meriem Laleg-Kirati is with INRIA Bordeaux Sud-Ouest, MAGIQUE-3D project team,
UFR Sciences, Bâtiment B1, Université de Pau et des Pays de l'Adour
BP 1155, 64013 Pau, France, (e-mail: Taous-Meriem.Laleg@inria.fr).}%
  \and
Claire~M\'edigue\thanks{Claire M\'edigue is with INRIA-Rocquencourt, B.P. 105, 78153 Le Chesnay cedex,
France, (e-mail: Claire.Medigue@inria.fr).}
\and
Yves Papelier\thanks{Yves Papelier is with EA 3544 EFM Hôpital Antoine Béclère 92141, Clamart,France, (e-mail: yves.papelier@kb.u-psud.fr)}
\and
Fran\c{c}ois Cottin\thanks{Fran\c{c}ois Cottin is with Unité de Biologie Intégrative des Adaptations à l'Exercice (INSERM 902 EA
3872, Genopole), 91000 Evry, France, (e-mail: francois.cottin@bp.univ-evry.fr)}
\and
Andry Van de Louw\thanks{Andry Van de Louw is with Intensive Care Unit, Centre Hospitalier Sud-Francilien,
91014 Evry, France, (e-mail: andry.vandelouw@ch-sud-francilien.fr)}%
}
\authorhead{Laleg \& M\'edigue \& Papelier \& Cottin \& Van de Louw }
\RRtitle{Validation d'une nouvelle méthode pour l'estimation du volume d'éjection systolique: comparaison avec le PiCCO}
\RRetitle{Validation of a New Method for Stroke Volume Variation Assessment: a Comparison with the PiCCO Technique}
\titlehead{Validation of a New Method for Stroke Volume Variation Assessment}

\RRresume{Cet article propose une nouvelle méthode pour l'estimation du volume d'ejection systolique par des mesures de pression artérielle. Le signal de  pression est reconstruit à l'aide d'une méthode d'analyse semi-classique permettant le calcul d'un paramètre, appelé le premier invariant systolique $INVS_1$. On montre que $INVS_1$ est linéairement relié au volume d'ejection systolique. Afin de valider cette approche, une comparaison statistique entre $INVS_1$ et le volume d'ejection systolique mesuré par la technique PiCCO a été effectuée pour un enregistrement de 15 minutes pour  21  patients mécaniquement ventilés et en soins intensifs. Pour  94\% de l'enregistrement complet, une forte corrélation a été estimée par une analyse cross-corrélation (coefficient moyen=0.9) et une regression linéaire (coefficient moyen =0.89). Une fois la relation linéaire vérifiée,  un test de  Bland-Altman a montré une bonne correspondance entre les deux approches et leur interchangeabilité. Pour les 6\% restant, $INVS_1$ et le volume d'éjection calculé par PiCCO  n'ont pas été corrélés, et cette différence, interprétée à l'aide de la pression moyenne, de la fréquence cardiaque et des résistances vasculaires périphériques a été en faveur de $INVS_1$.
}

\RRabstract{This paper proposes a novel, simple and minimally invasive method for stroke volume variation assessment using arterial blood
pressure measurements. The arterial blood pressure signal is reconstructed using a semi-classical signal analysis method allowing the
computation of a parameter, called the first systolic invariant $INVS_1$. We show that $INVS_1$ is linearly related to stroke volume.
To validate this approach, a  statistical comparaison between $INVS_1$ and stroke volume measured with the PiCCO technique  was performed during
a 15-mn recording in  21  mechanically ventilated patients in intensive care. In 94\% of the whole recordings, a strong correlation was
estimated  by  cross-correlation analysis (mean coefficient=0.9) and linear regression (mean coefficient=0.89). Once the linear relation
had been verified,  a  Bland-Altman test showed the very good agreement between the two approaches and their interchangeability.
For the remaining  6\%, $INVS_1$ and the PiCCO stroke volume were not correlated at all, and  this discrepancy,
interpreted with the help of mean pressure, heart rate and peripheral vascular resistances,  was in favor of  $INVS_1$.}
\RRmotcle{Pression artérielle, premier invariant systolique,  PiCCO, analyse semi-classique du signal, variations du volume d'éjection}
\RRkeyword{Arterial blood pressure, first systolic invariant, PiCCO, semi-classical signal analysis, stroke volume variation}
 \RRprojet{SISYPHE}  
\RRdomaine{5} 
\RRtheme{Observation, modélisation et commande pour le vivant}
 \URRocq 

\begin{document}
 \RRNo{7172}
\makeRR   
\tableofcontents
\section{Introduction}
Hemodynamic monitoring is crucial for critical care patient management. A recent international consensus conference recommended against the routine use of static preloaded measurements alone to predict fluid responsiveness \cite{AnLeAnChHuMaMeMoPuStTo:07}, and dynamic assessment now seems more useful. Several studies have documented the ability of respiratory stroke volume variation ($SVV$) to predict the fluid responsiveness in hemodynamically compromised patients \cite{CaMuDeBoSiHeLe:09}-\cite{HuFuHuKaChHsTs:08}.
Measuring respiratory $SVV$ requires a continuous monitoring of stroke volume ($SV$) which can be obtained using invasive or non-invasive methods. Current  invasive methods used have the disadvantage of requiring the insertion of a central venous catheter and the calibration of the cardiac output measure with a cold isotonic sodium chloride bolus (PiCCO technology) \cite{BeWoToPaStHeSc:04} or a lithium chloride bolus (LiDCO technology) \cite{CeDaGrRh:09}. An alternative method, which does not require venous catheter insertion or calibration  has been proposed (Flotrac Vigileo) \cite{BiNoRoQuReSz:09}, but several clinical studies have pointed out its poor agreement with reference techniques \cite{DeGeVaJa:09}, \cite{LaKaMaChPeKaFlHe:09}. Esophageal echo-doppler is the main non-invasive method, calculating aortic blood flow from the echo-derived aortic diameter and the doppler-derived aortic blood velocity \cite{DaSi:04}. Nevertheless, this technique has potential contraindications, such as esophageal varices or esophageal surgery, and several limitations: for instance, it measures the blood flow in the descending aorta and not the whole cardiac output. Moreover, the precision of the measurement depends on accurate probe positioning, which is not always easy to obtain \cite{BeTiFoMa:98}.
Thus, each of the above methods has its own drawbacks, and there is still a need for an easily applicable, minimally invasive, accurate and affordable  method to estimate $SVV$.

Due to the fact that Arterial Blood Pressure (ABP) can be measured using minimally invasive or noninvasive methods, the idea of estimating $SV$ from ABP has captured scientists for a long time.  Thus, many methods have been developed and whose objective is to find a relation between one or several parameters characterizing the shape of the pressure and $SV$ or cardiac output ($CO$), see for instance \cite{BoGiBeWo:76}, \cite{KoShMc:70}, \cite{LiLi:01}, \cite{MuReHoMaCo:06}  and the references quoted there. These methods, which are based  on some models of systemic circulation, are called pulse contour methods. A comparison between some of the pulse contour methods has been proposed in \cite{AlBrSaBrHa:72}, \cite{StMcCoGr:73}, \cite{SuReSaMa:05}, \cite{YuDiLiSaSpToPo:98}. The simplest model supposes a proportionality between  $CO$ and the Mean Arterial Pressure ($MAP$). Other approaches, based on windkessel models, link $SV$ to different lumped parameters such as pulse pressure, the systolic and diastolic pressures \cite{BoGiBeWo:76}. However these approaches consider the arterial system as a lumped system which appears not sufficiently accurate. So, other methods resulting from distributed arterial models use the pressure area so that $SV$ is often supposed to be proportional to the area under the systolic part of the pressure curve. Corrected versions of this relation have been also proposed \cite{KoShMc:70}. However, this approach requires detecting the end of the systole
which is completely nontrivial, particularly in peripheral ABP waveforms.
Moreover, approaches taking into account the nonlinear aspects of the arterial system have been proposed, for example modelflow \cite{WeJaSeSc:93}, but some studies have revealed the poor efficiency of this method in a number of cases \cite{ReAeVe:02}.

In this paper we introduce a novel technique for $SVV$ assessment using ABP measurements. This method is based on the analysis of ABP with a new signal analysis method that was recently proposed in \cite{Laleg:08}, and called Semi-Classical Signal Analysis (SCSA). The new spectral parameters provided by SCSA, eigenvalues and invariants, have already given promising results in some other applications, as summarized in the following.

On the one hand, we assessed their ability to discriminate between different situations. In the first situation, nine heart failure subjects were compared to nine healthy subjects.  In the second situation, eight highly fit triathletes were compared before and after training. SCSA parameters always provided  more significant results than classical parameters, regarding temporal as well as spectral parameters (\cite{LaMeCoSo:07}, \cite{Laleg:08}).
On the other hand, we tested the ability of the invariants to represent physiological parameters of great interest, particularly $SVV$,
in two well-known conditions:  the head-up 60 degrees tilt-test and the handgrip-test \cite{Laleg:08}.
Let us focus  on the first invariants. The first global invariant ($INV_1$) is, by definition, the mean value of the ABP signal,
which is a standard parameter in clinical practice. The  first systolic ($INVS_1$) and diastolic ($INVD_1$) invariants are less obvious. They result from the decomposition of the pressure into its systolic and diastolic parts. In particular, $INVS_1$ corresponds to the integral of the estimated systolic pressure with SCSA. Referring to the pulse contour method stating that the area under the systolic part of the pressure curve is proportional to $SV$ as described above, one can show  that $INVS_1$ variations give information on $SVV$.



We study in this paper the correlation between $INVS_1$ and measured $SVV$ using a reference method; the PiCCO technique. The PiCCO technique uses the pulse contour method with a calibration by a transpulmonary thermodilution and is considered a reliable technique.
In what follows, we present the experimental protocol and recall some basic aspects of the SCSA method. We introduce $INVS_1$ and its relation to $SVV$. Then,  we present  statistical results  on 21 patients' recordings.

\section{Materials and Methods}
This prospective study was conducted 
in the 16-bed medical-surgical intensive care unit (ICU) of the Sud-Francilien General Hospital (Evry, France).

\subsection{Patients}
\begin{itemize}
\item  \textbf{\emph{Inclusion criterion:}} all mechanically ventilated patients whose cardiac output was continuously monitored with a transpulmonary thermodilution catheter (PiCCO, Pulsion Medical Systems, Munich, Germany) were included, except those satisfying the following excluding criteria. PiCCO is routinely used in this unit to monitor hemodynamically compromised patients.

\item \textbf{\emph{Exclusion criteria:}} patients presenting cardiac arrhythmias or breathing spontaneously were excluded because the SVV is not applicable for such patients.

\item \textbf{\emph{Protocol:}} all patients were sedated with midazolam and fentanyl in dosages that were titrated to achieve full adaptation to the ventilator. Ventilator settings were as follows: volume assist-control mode; tidal volume (Vt), $6 ml/kg$ ideal body weight; breathing rate, 20 cycles/minute; inspiratory/expiratory ratio, $\dfrac{1}{2}$; and $FiO2$ adjusted to maintain transcutaneous oxygen saturation in blood 94\%. Positive end-expiratory pressure (PEEP) was set at $5 cm$ $H2O$ but some hypoxemic patients required an increase in PEEP to 10 cm H2O during the data acquisition, to improve arterial oxygenation. The increase in PEEP was left to the discretion of the attending physician, as well as the adaptation of vasoactive drugs dosages, adjusted to maintain an adequate circulatory status during the protocol.
\end{itemize}

\subsection{Data acquisition}
One-lead electrocardiogram, arterial pressure, and respiratory flow signals were recorded during a 15-min period using a Biopac 100 system (Biopac systems, Goleta, CA, USA). All data were sampled at $1000 Hz$ and stored on a hard disk. Cardiac output was calibrated just before the data acquisition with a cold isotonic sodium chloride bolus of 20 ml. Then, $CO$ and peripheral vascular resistances ($PVR$) were delivered every 30 seconds during the 15-min period.

\subsection{Signal analysis}

Signal processing was performed using the Scilab and Matlab environments at the French National institute for Research in Computer Science and Control (INRIA-Sisyphe team).\\

\subsubsection*{A Semi-Classical Signal Analysis method}

In this section, we introduce the SCSA technique and some results of its application to ABP analysis. We also show the relation between $INVS_1$ and $SVV$.

\paragraph{\textbf{The SCSA principle}}

Let $y: t\longmapsto y(t) $ be a real valued function representing the signal to be
analyzed such that:
\begin{eqnarray}\label{hypotheses}
y \in L_1^1(\mathbb{R}), \: y(t)\geq 0,\:  \forall t\in
\mathbb{R},\nonumber\\
\frac{\partial^m y}{\partial x^m}\in
L^1(\mathbb{R}), \: m=1,2,
\end{eqnarray}
with,
\begin{equation}\label{condpotentiel} L_1^1(\mathbb{R})=
\{V | \int_{-\infty}^{+\infty}{|V(t)|(1+|t|) dt}<\infty\}.
\end{equation}

The main idea in the SCSA consists in interpreting the signal $y$ as a multiplication operator, $\phi \rightarrow y.\phi$, on some function space.
 Then, instead of the standard Fourier Transform,  we use the spectrum of a regularized version of this operator, known as the Schrödinger operator in $L^2(\mathbb{R})$, for the analysis of $y$:
\begin{equation}\label{schrchap2}
H(h ; y) = -h^2\frac{d^2}{dt^2} - y,
 \end{equation}
 for a small $h > 0$.  The SCSA method is better suited to the analysis of some pulse shaped signals than the Fourier Transform \cite{Laleg:08}.

In this approach, the signal is a potential of the Schrödinger operator $H(h ; y)$. We are interested in the spectral problem of this operator  which is given by:
\begin{equation}\label{sch}
-h^2\frac{d^2 \psi}{dt^2} - y \psi = \lambda \psi,\quad t\in\mathbb{R},
\end{equation}
where $\lambda$,  $\lambda \in \mathbb{R}$ and $\psi$, $\psi\in H^2(\mathbb{R})$ \footnote{$H^2(\mathbb{R})$ denotes the Sobolev space of order 2} are respectively the  eigenvalues of $H(h; y)$ and the associated eigenfunctions.  Under equation  (\ref{hypotheses}),
 the spectrum  of $H(h; y)$ consists of:
\begin{itemize}
    \item a continuous spectrum $\lambda \geq 0$,
    \item a discrete spectrum composed of negative eigenvalues. There is a non-zero, finite number $N_h$ of  negative eigenvalues of the
operator $H(h; y)$. We put $\lambda = - \kappa_{nh}^2$ with $\kappa_{nh} > 0$ and $\kappa_{1h}> \kappa_{2h}> \cdots > \kappa_{nh}$,
$n=1,\cdots,N_h$. Let $\psi_{nh}$, $n=1,\cdots, N_h$ be
the associated $L^2$-normalized eigenfunctions \cite{Laleg:08}.\\
\end{itemize}

The SCSA technique consists in reconstructing the signal $y$ with the discrete spectrum of $H(h;y)$ using the following formula:
\begin{equation}\label{formule introduction}
y_h(t)= 4 h \sum_{n=1}^{N_h}{\kappa_{nh
}\psi_{nh}^2(t)}, \quad \quad t\in \mathbb{R}.
\end{equation}

Here, the parameter $h$ plays an important role. As $h$ decreases, the approximation of the signal improves.
However,  as $h$ decreases, the number of negative eigenvalues $N_h$ increases and hence the time required to perform the computation increases. So, in practice, what we are looking for is a value of $h$ that provides a sufficiently small estimation error with a reduced number of negative eigenvalues.
We summarize the main steps for reconstructing a signal with the SCSA as follows \cite{Laleg:08}:
\begin{enumerate}
\item Interpret the signal to be analyzed  $y$ as a potential of the Schrödinger operator $H(h;y)$ (\ref{schrchap2}) ;

\item compute the negative eigenvalues and the associated $L^2$-normalized eigenfunctions of $H(h;y)$ ;

\item  compute $y_h$ according to equation (\ref{formule introduction}) ;

\item look for a value of $h$ to obtain a good approximation with a small number of negative eigenvalues.\\

\end{enumerate}

\paragraph{\textbf{ABP analysis with the SCSA}}

Now, we introduce some results on the application of the SCSA to ABP analysis.
We denote by $P$ the ABP signal and $\hat{P}$ its estimation with the SCSA such that:
\begin{equation}\label{pestime}
    \hat{P}(t)=4h\sum_{n=1}^{N_h}{\kappa_{nh}\psi_{nh}^2(t)},
\end{equation}
where $-\kappa_{nh}^2$, $n=1,\cdots,N_h$ are the $N_h$
negative eigenvalues of the Schr\"{o}dinger  operator $H(h;P)$ and  $\psi_{nh}$ the associated $L^2-$normalized
eigenfunctions.

The ABP signal was estimated for several values of the parameter $h$ and hence $N_h$.  Fig.\ref{pression
doigt2} illustrates measured and estimated pressures for one beat of an ABP signal and the estimated error with $N_h=9$. Signals measured at the aorta (invasively) and at the finger (non invasively) respectively were considered.  We point out that $5$ to $9$ negative eigenvalues are sufficient for a good estimation of an ABP beat \cite{LaCrPaSo:07}, \cite{LaCrSo:07}.

\begin{figure}[htbp]
\begin{center}
\subfigure{\epsfig{figure=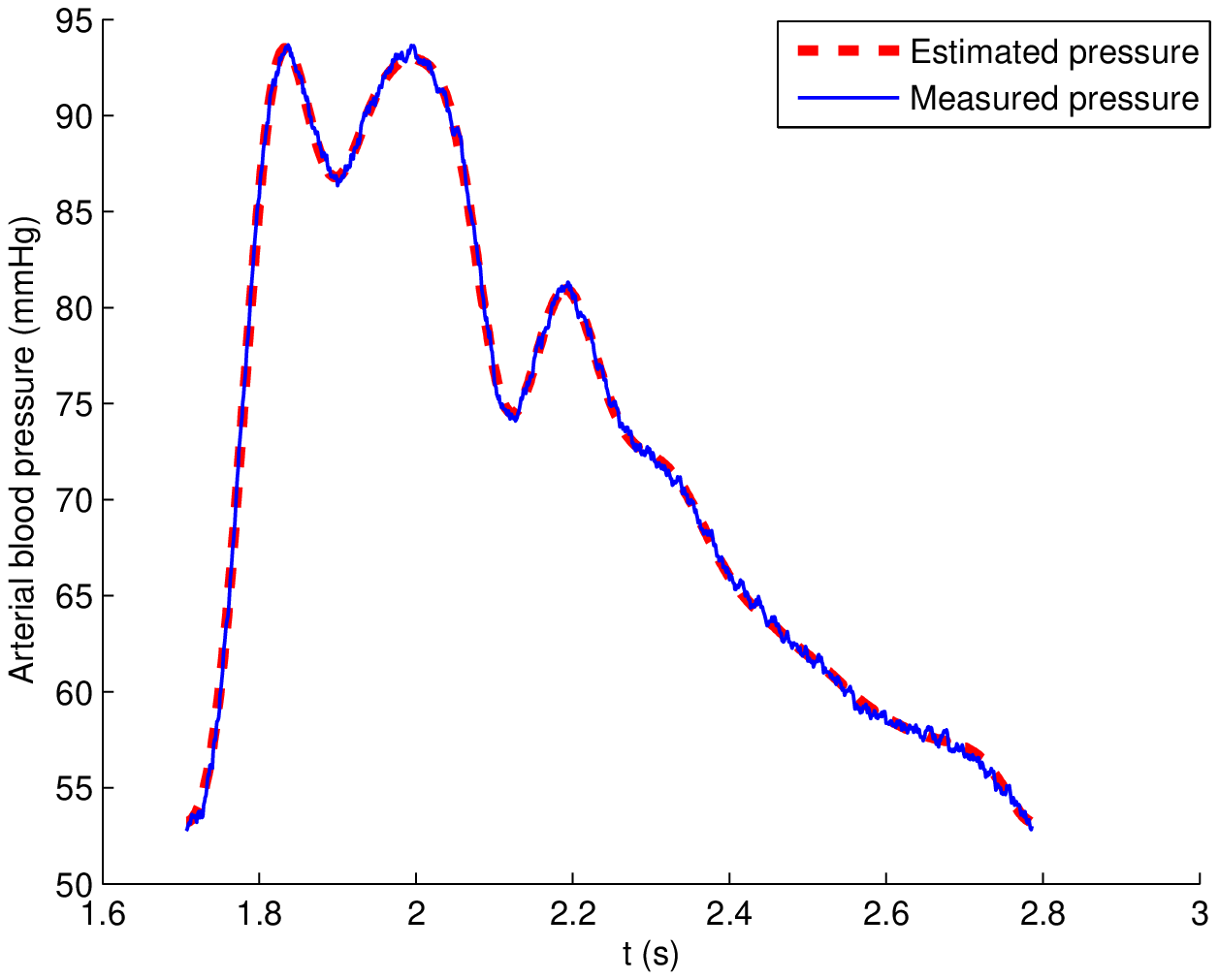,width=6cm}}
\subfigure{\epsfig{figure=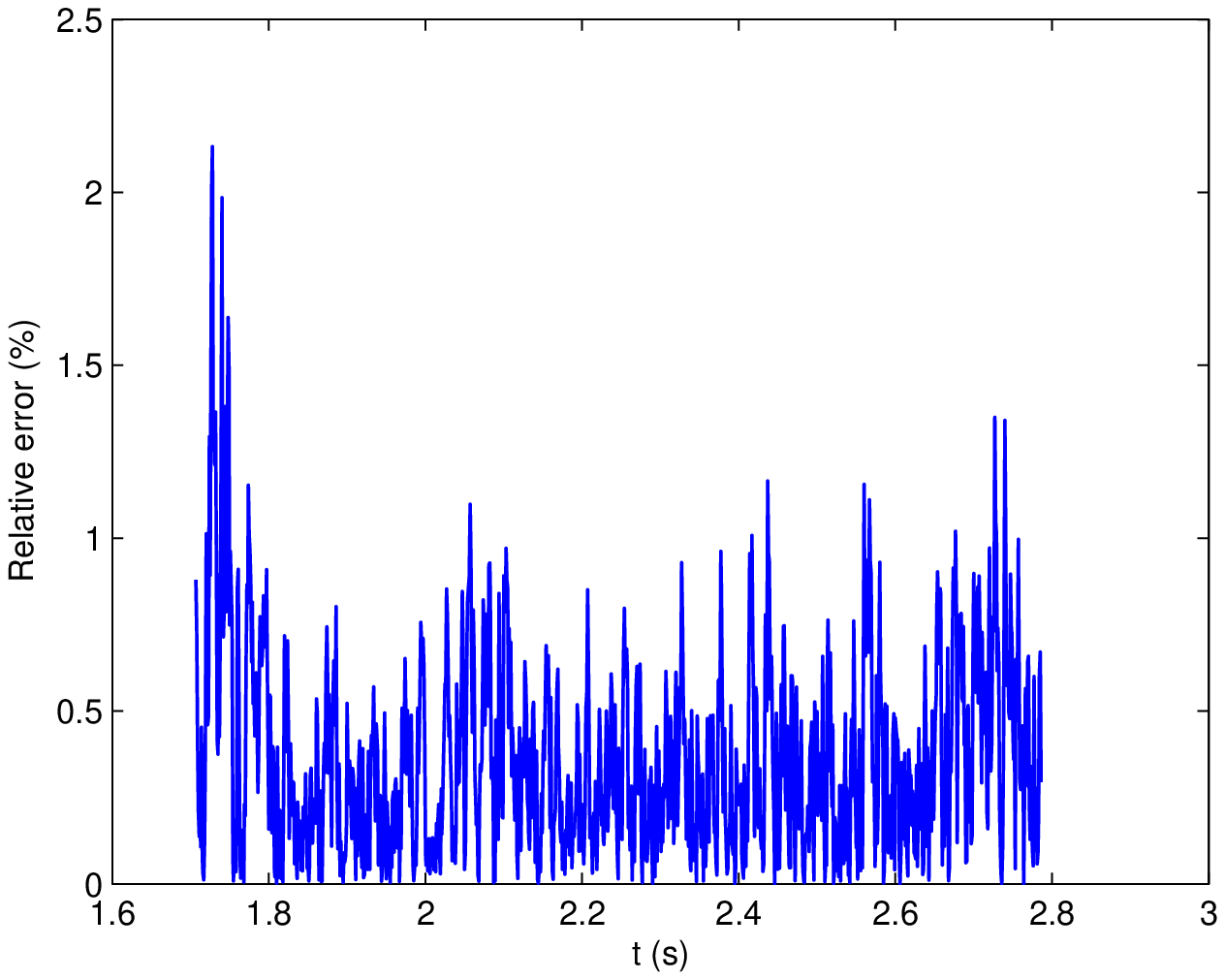,width=6cm}}
\begin{center}
{(a) Aorta}\end{center}
\subfigure{\epsfig{figure=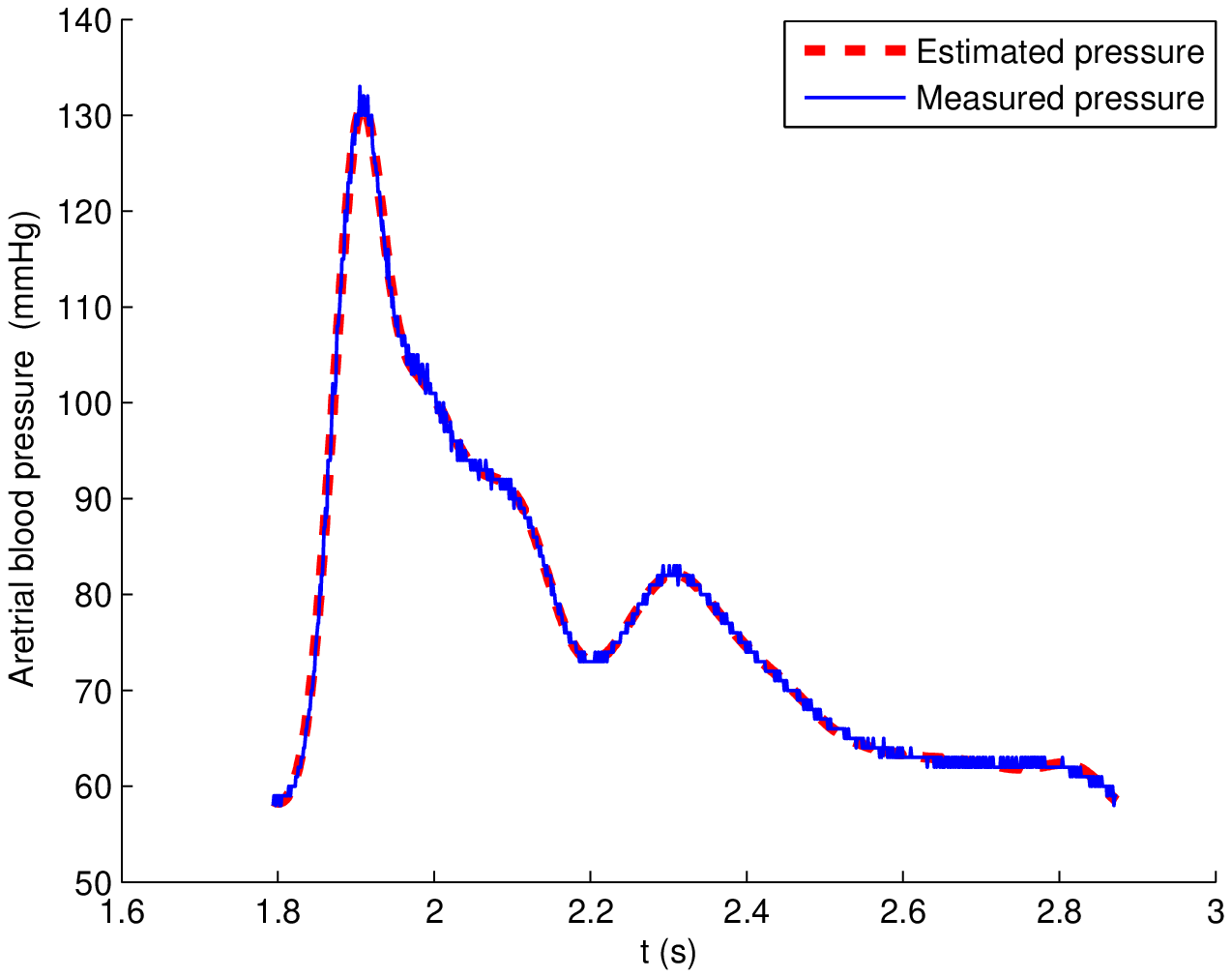,width=6cm}}
\subfigure{\epsfig{figure=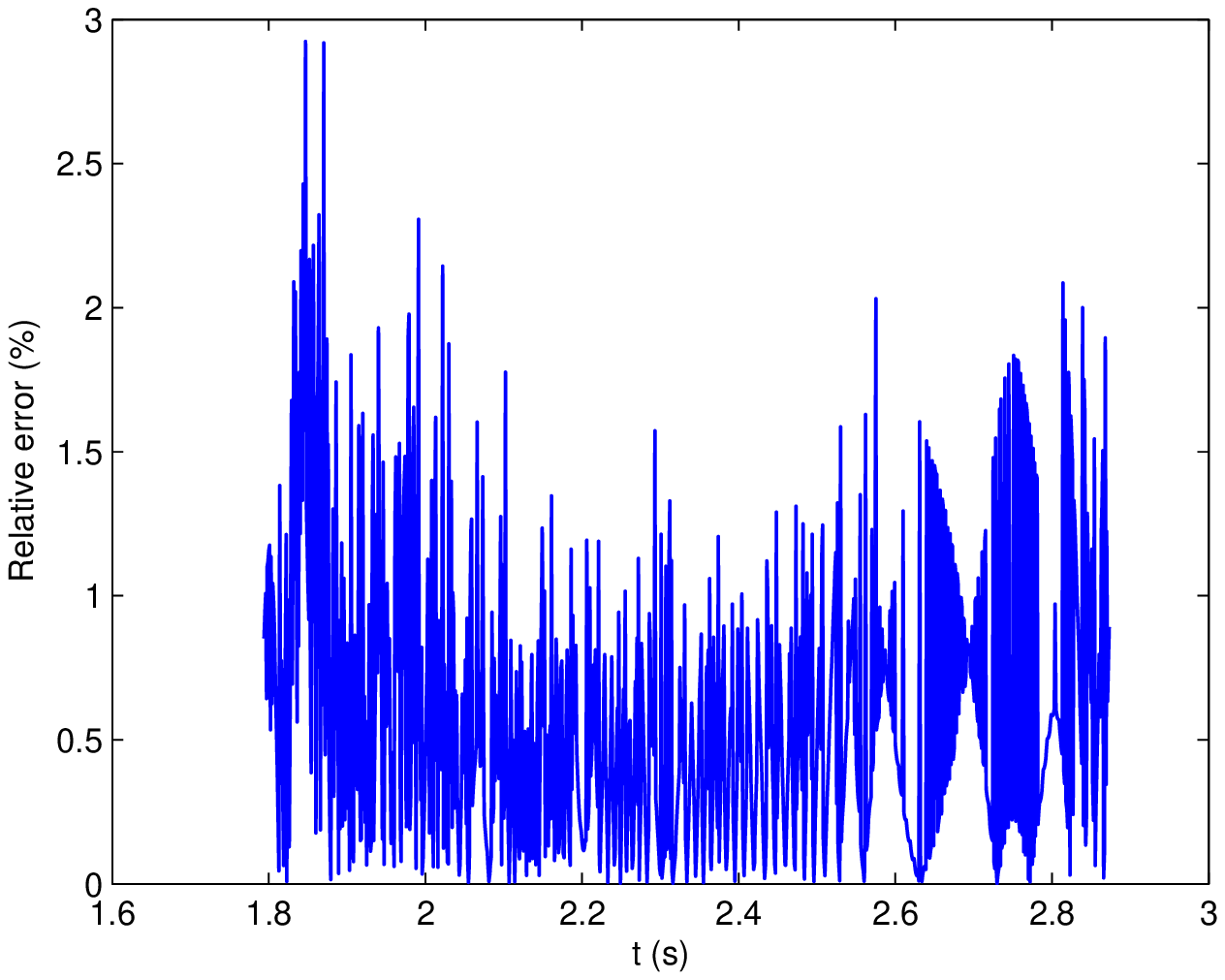,width=6cm}}
\begin{center}
{(b) Finger}\end{center}
 \caption{Estimation of the pressure at the aorta and the finger level with the SCSA and $N_\chi=9$. On the left, the estimated and measured pressures.  On the right, the relative error}\label{pression doigt2}
\end{center}
\end{figure}

One application of the SCSA to ABP signals consists in decomposing the signal into its systolic and diastolic parts. This application was inspired by a reduced model of ABP based on solitons solutions of a Korteweg-de Vries (KdV) equation \footnote{Solitons are solutions of some non-linear partial derivative equations like the KdV equation}  proposed in \cite{CrSo:07}, \cite{LaCrSo:07J}. As described in  \cite{LaCrPaSo:07}, \cite{Laleg:08}, the idea consists in decomposing  (\ref{pestime}) into two partial sums: the first one, composed of the $N_s$ ($N_s=1, 2, 3$ in general) largest  $\kappa_{nh}$
and the second composed of the remaining components. Then, the first partial sum represents rapid phenomena
that predominate during the systolic phase and the second one describes slow phenomena of the diastolic phase. We denote by $\hat{P}_s$ and
$\hat{P}_d$ the systolic pressure and the diastolic pressure respectively estimated with the SCSA. Then we have:
\begin{eqnarray}\label{pression systolique}
    \hat{P}_s(t) &=  & 4 h\sum_{n=1}^{N_{s}}{\kappa_{nh}\psi_{nh}^2(t)},
\end{eqnarray}
\begin{eqnarray}
     \hat{P}_d(t) & = &4 h \sum_{n=N_{s}+1}^{N_h}{\kappa_{nh}\psi_{nh}^2(t)}.
\end{eqnarray}

\begin{figure}[htbp]
 \begin{center}
\subfigure[]{\epsfig{figure=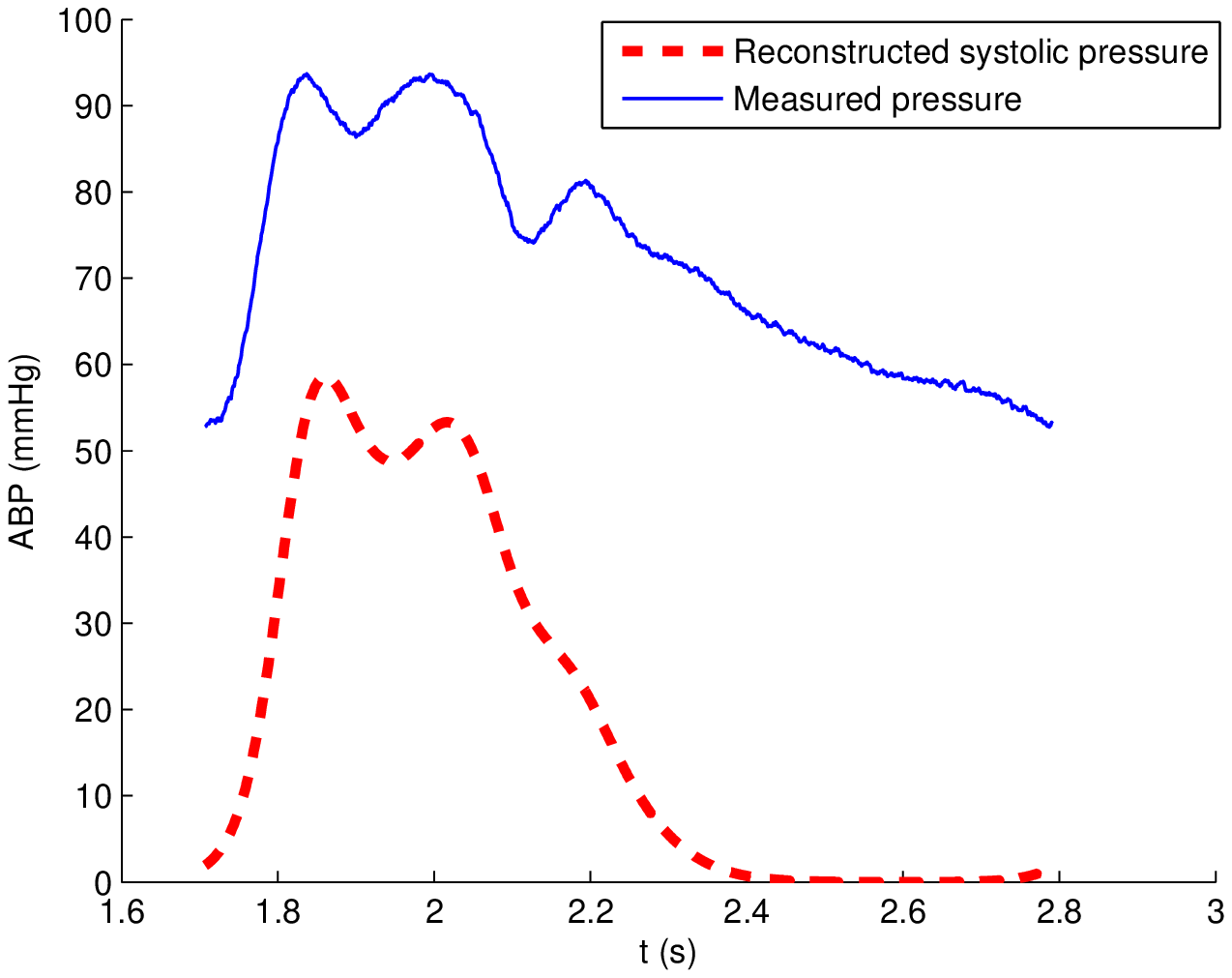,width=6cm}}
\subfigure[]{\epsfig{figure=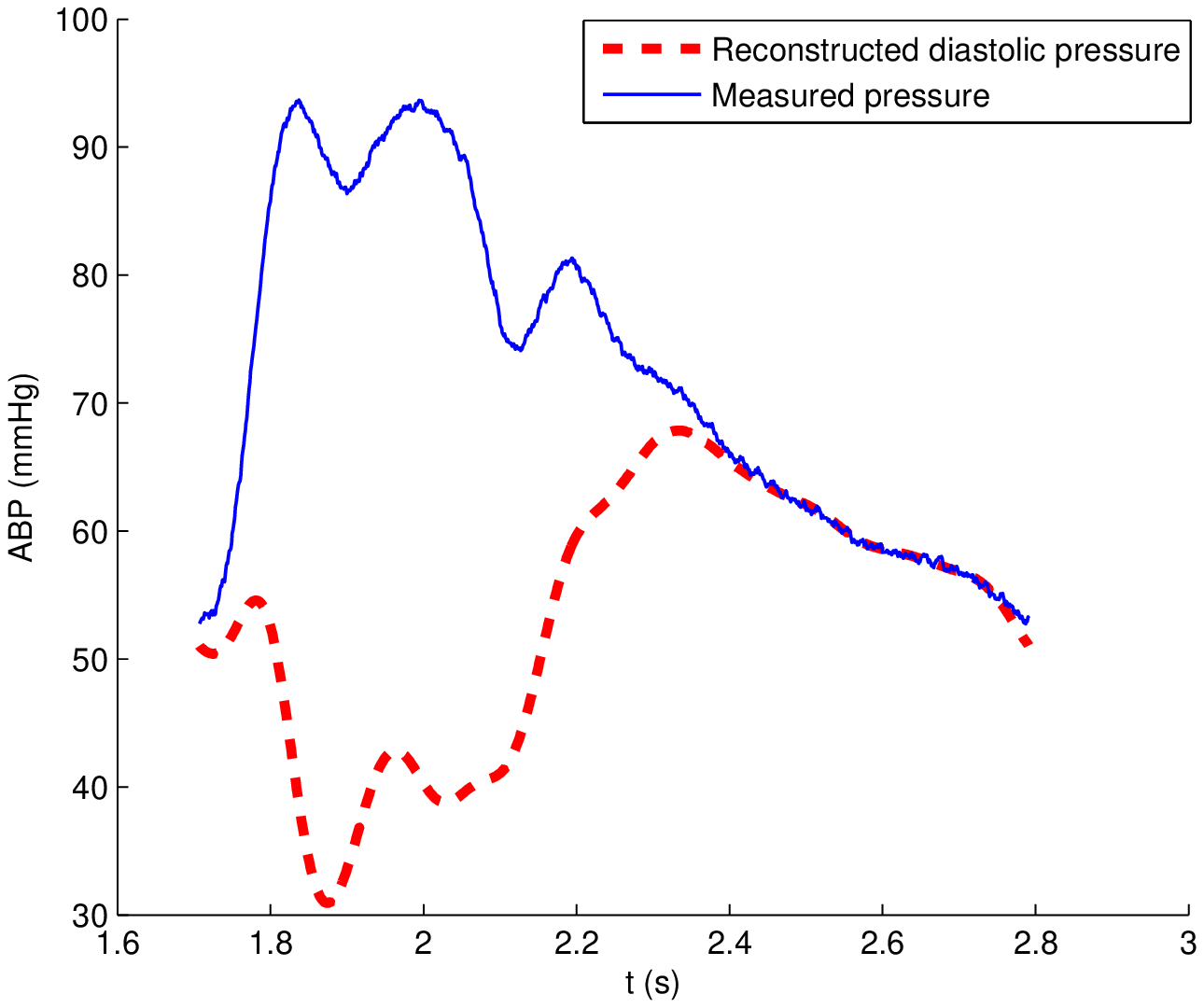,width=6cm}}
   \caption{(a) Estimated systolic pressure, (b) Estimated diastolic pressure}\label{decomposition}
\end{center}
\end{figure}

Fig.\ref{decomposition}  shows  measured pressure and estimated systolic and diastolic pressures respectively. We notice that $\hat{P}_s$ and $\hat{P}_d$ are respectively localized during the systole and the diastole.\\

\paragraph{\textbf{SCSA parameters}}

As seen previously, the SCSA technique provides a new description of the
ABP signal with some spectral parameters which are the negative eigenvalues and the so called invariants \footnote{We call these parameters invariants because they are related to the Korteweg-de Vries invariants in time}. The latter consist in some momentums of the $\kappa_{nh}$, $n=1,\cdots,N_h$. So we define the first two global invariants by:

\begin{equation}
INV_1=4h \sum_{n=1}^{N_h}{\kappa_{nh}},
\quad
INV_2=\frac{16}{3}h\sum_{n=1}^{N_h}{\kappa_{nh}^3}.
\end{equation}

Systolic ($INVS_{1,2}$) and diastolic ($INVD_{1,2}$) invariants are deduced from the decomposition of the pressure into its systolic and diastolic parts and are then given by:
\begin{equation}\label{invariants systoliques}
INVS_1=4h \sum_{n=1}^{N_s}{\kappa_{nh}},
\end{equation}
\begin{equation}
INVS_2=\frac{16}{3}h\sum_{n=1}^{N_s}{\kappa_{nh}^3},
\end{equation}
\begin{equation}
INVD_1=4h \sum_{n=N_s+1}^{N_h}{\kappa_{nh}},
\end{equation}
\begin{equation}
INVD_2=\frac{16}{3}h\sum_{n=N_s+1}^{N_h}{\kappa_{nh}^3}.
\end{equation}

\paragraph{\textbf{$\boldsymbol{INVS_1}$ for $\boldsymbol{SVV}$ estimation}}

\begin{figure}
\begin{center}
  \includegraphics[width=8cm]{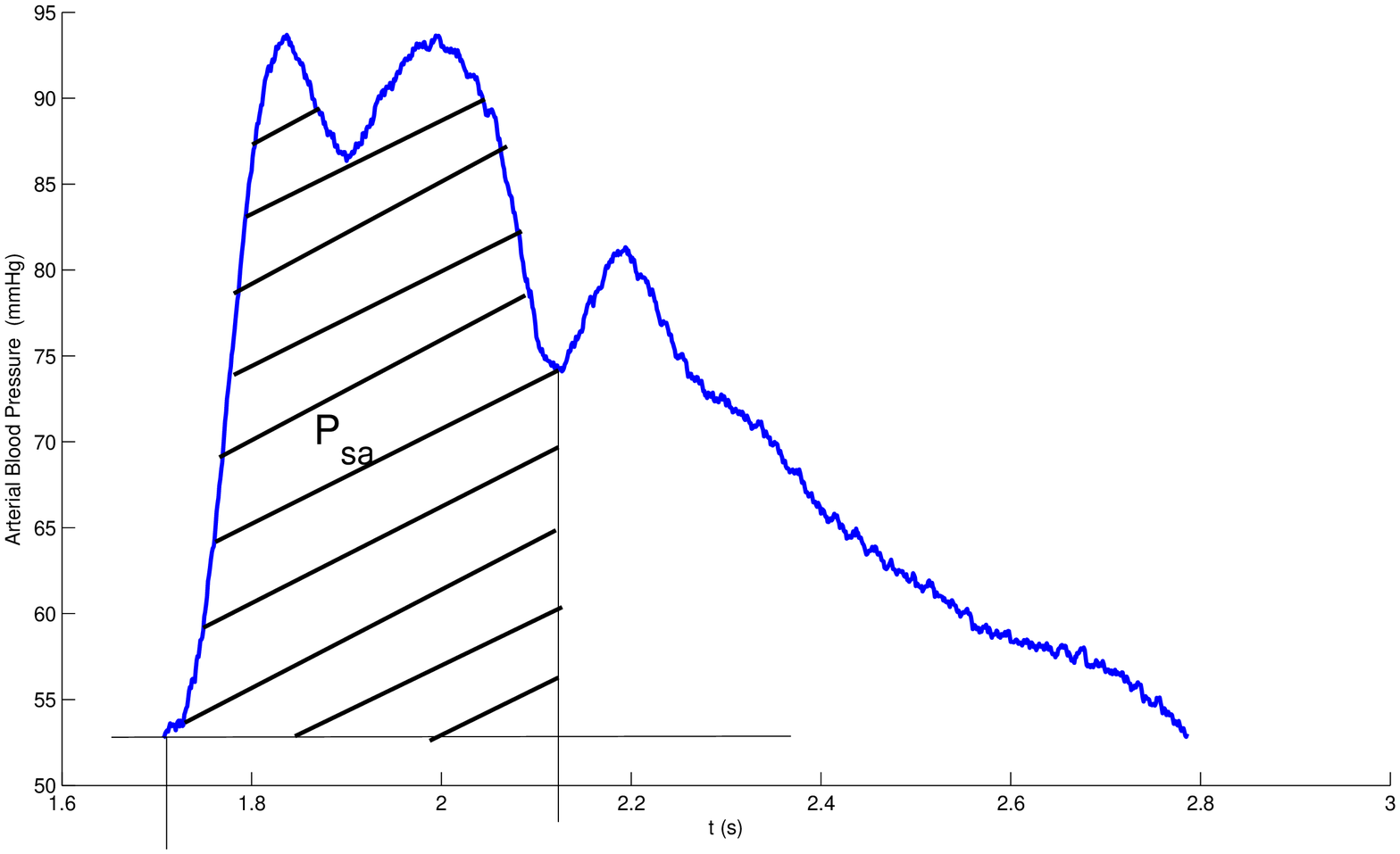}\\
  \caption{Area under the systolic part of the pressure curve used to estimate $SV$}\label{pulse contour}
\end{center}
\end{figure}

We will see here how $INVS_1$ is related to  $SV$. For this purpose, we recall one approach of the pulse contour methods that supposes proportionality between $SV$ and the area under the systolic part of the pressure curve as described in the introduction. We denote this area by $P_{sa}$ (see fig.\ref{pulse contour}). So we have:
\begin{equation}
SV_{PC} = k P_{sa},
\end{equation}
where $k$ is a positive real and $SV_{PC}$ is the stroke volume estimated with the pulse contour method.

Referring to (\ref{pression systolique}) and (\ref{invariants systoliques}), and remembering that $\psi_{nh}$ are $L^2$-normalized, we have:
\begin{equation}
    INVS_1= \int_{-\infty}^{+\infty} {\hat{P}_s(t)dt}.
\end{equation}
So, $INVS_1$ refers to the area under the systolic curve $\hat{P}_s$.
Thus, one can remark that both $P_{sa}$ and $INVS_1$ describe the area under the systolic pressure but they may not be equal because the detection of the end systole in the two cases is not the same (see fig.\ref{decomposition}.a and fig.\ref{pulse contour}). Indeed, while $P_{sa}$ is computed by detecting the dicrotic notch which is completely non trivial in peripheral ABP waves, $INVS_1$ results from a nonlinear model of ABP based on solitons that considers  the propagation of the pulse wave \cite{CrSo:07}, \cite{LaCrSo:07J} as was described in section II.C.1.b. We can write the following relation between the two areas:
\begin{equation}
    INVS_1 = P_{sa} + b,
\end{equation}
where $b \in \mathbb{R}$ represents the difference between the two areas. Then, we get:
\begin{equation}\label{regression lineaire}
    INVS_1= a SV_{PC} + b.
\end{equation}
with $a=\dfrac{1}{k}$.  Hence, $INVS_1$ and $SV_{PC}$ are linearly related.\\

\subsubsection*{Resulting time series used to compare $\boldsymbol{INVS_1}$ to  PiCCO stroke volume}

On top of $INVS_1$, two vascular time series were analyzed: the heart rate ($HR$) was computed from the pulse interval ($PI$), which is the distance between two systolic occurrences; $MAP$ was calculated from the systolic and diastolic values. All data were resampled at 4 Hz, by the interpolation of a third order spline function to obtain equidistant data and to guarantee their synchronization. They were then averaged over 15 seconds and delivered every 30 seconds, like the PiCCO protocol. PiCCO cardiac output was divided by $HR$ to give a stroke volume ($SV_{PiCCO}$).
According to the relation between $INVS_1$ and $SV$, $INVS_1$  was subsequently called  $SV_{SCSA}$ when it was estimated with the linear equation (\ref{regression lineaire}):  $SV_{SCSA} = a SV_{PiCCO} + b$. Thus, in addition to the $SV_{PiCCO}$  and $ SV_{SCSA}$, temporal relations with $HR$, $MAP$ and $PVR$ time series were analyzed to help interpret $SV$ behavior in case of a divergence between $SV_{PiCCO}$ and $SV_{SCSA}$.

\subsection{Statistical analysis}

\begin{enumerate}
\item  \textbf{Cross-correlation analysis}.\\

The cross-correlation analyzes the temporal similarity between two time series by estimating the correlation
between one time series at time t and the other at time $t\pm x$ lags (in samples) \cite{ShSt:00}.
A cross-correlation was performed between $SV_{PiCCO}$ and $INVS_1$ time series averaged every 30 seconds.
Cross-correlation coefficients were computed using the Matlab xcorr function (The MathWorks, Inc) after subtracting the means from the time series.
Cross-correlation coefficients were computed for all lags $(-30;+30)$, each lag corresponding to a 30-second interval. The correlation coefficients
of the unlagged data stand at the midpoint (lag 0). A 5\% level of probability for the correlation coefficients was considered significant (Bravais-Pearson table). An average estimate of the correlation over all the subjects was allowed after homogeneity tests (non significant z-test and  Jarque-Bera test) \cite{JaBe:87}. The individual correlation coefficients were averaged for each lag.\\

\item  \textbf{Linear regression}.\\

According to the proportional relation between $SV_{PiCCO}$ and $INVS_1$ (equation (\ref{regression lineaire})), a linear regression analysis was applied, using SigmaStat. This analysis provides the Pearson $R$ coefficient, which measures the degree of linear correlation between the two estimates, and the parameters of the linear equation (\ref{regression lineaire}), $a$ and $b$. So, they allow an estimation of the stroke volume $SV_{SCSA}$ using (\ref{regression lineaire}). This transformation is required before using the Bland-Altman test.\\

\item  \textbf{Bland-Altman method}.\\

Unlike the first two approaches which are not affected by differences in units or the nature of results, the Bland-Altman method
analyzes the agreement between two estimates of the same variable \cite{BlAl:86}. Thus, we used $SV_{SCSA}$ instead of $INVS_1$, and moreover, $SV_{SCSA}$ variations to compare them to $SV_{PiCCO}$ variations ($\Delta:= SVV_{PiCCO} - SVV_{SCSA}$ with $SVV_{PiCCO} = SV_{PiCCO}(n) - SV_{PiCCO}(n-1)$ and  $SVV_{SCSA} = SV_{SCSA}(n) - SV_{SCSA}(n-1)$). The mean difference between $SVV_{PiCCO}$ and $SVV_{SCSA}$ (mean $\Delta$) is plotted against the average of the two volume variations. Mean $\Delta$ which represents the bias between the two methods, and the 95\% confidence interval ($[CI_{inf} \Delta \:\:\ CI_{sup}\Delta]$) gives the variation of the values of one method compared to the other.

\end{enumerate}

\section{Results}

\subsection{Patients}

The 21 patients recordings were analyzed over 900 seconds, representing about 30 averaged values, except one recording, analyzable only for the first 16 averaged values. In order to illustrate the main results, we choose the first three subjects in Table \ref{table1}, representing various conditions: subject one was submitted to PEEP changes (fig.\ref{sujet1}), subject two was submitted to noradrenaline dose changes (fig.\ref{sujet2}), subject three had no change in ventilatory parameters nor in drugs (fig.\ref{sujet3}).
\begin{table}[tb]
\begin{center}
\caption{Cross-correlation and linear regression coefficients between $SV_{PiCCO}$ and $INVS_1$} \label{table1}
\begin{tabular}{|c|c|c|c|} 
  \hline
 Subject & Number of & Cross & Linear \\
 &values&correlation&regression\\
 \hline
 1 & 30 &  0.99 & 0.98 \\
 \hline
2 &  30 &  0.97 &  0.97 \\
\hline
3 & 30 & 0.97 & 0.97  \\
\hline
4 & 30 & 0.96 & 0.96 \\
\hline
5& 30 & 0.96 & 0.96 \\
\hline
6 & 30& 0.95 & 0.95 \\
\hline
7 & 30& 0.95 & 0.95 \\
\hline
8 & 30 & 0.93  & 0.94 \\
\hline
9 & 30 & 0.91  & 0.91  \\
\hline
10  & 30 & 0.91 & 0.89  \\
\hline
11 & 30 & 0.90  & 0.89  \\
\hline
12 & 30  & 0.90  & 0.90  \\
\hline
13 & 30 & 0.86  & 0.85 \\
\hline
14  & 30 & 0.85  & 0.77  \\
\hline
15 & 30& 0.85 & 0.86  \\
\hline
16  & 30  &  0.85  & 0.85   \\
\hline
17  & 21  & 0.84  & 0.87  \\
\hline
18  & 30  & 0.82 & 0.83  \\
\hline
19  & 16  & 0.82 & 0.82  \\
\hline
20 & 30  & 0.77  & 0.77  \\
\hline
\end{tabular}
\end{center}
\small{The subjects (col. 1) are listed in the decreasing order of cross-correlation coefficients. Col. 2 indicates the number of measurements delivered every 30-second per subject, according to the PiCCO  protocol and representing a 900 seconds analysis. Col. 3 and 4 stand  for  coefficients of cross-correlation (mean: $0.90 \pm 0.01$ at lag 0) and linear regression  (mean: $0.89 \pm 0.01$). Subject 21 and the last third of Subject 17, whose $INVS_1$ and $SV_{PiCCO}$ were not correlated at all (coefficient $<0.1$), were  discarded from the table. The last part of Subject 19 was discarded before analysis because of a  PiCCO dysfunction.}
\end{table}

\begin{table}[tb]
\begin{center}
\caption{Bland-Altman test results} \label{table2}
\begin{tabular}{|c|c|c|c|c|}
  \hline
Subject & $Min \Delta$  & $CI inf \Delta$ & $Max \Delta$ & $CI sup \Delta$ \\
\hline
1 & -.0056  & -.0074  &  .0048  &  .0077   \\
\hline
2  & -.0026  & -.0037  &  .0034  &  .0039  \\
\hline
3  & -.0030  & -.0036  & .0022  &  .0037    \\
\hline
4  & -.0043  & -.0057  &  .0043  &  .0056   \\
\hline
5  & -.0034  & -.0039  &  .0044  &   .0040  \\
\hline
6  & -.0048  & -.0052  &  .0045  &  .0053  \\
\hline
7 &  -.0059  & -.0072   &  .0043 & .0072  \\
\hline
8 &  .0005  & -.0009  &  .0009  &  .0009 \\
\hline
9 & -.0037 & -.0051 &  .0046 &   .0050 \\
\hline
10  & -.0047  & -.0055  &  .0042  &  .0057   \\
\hline
 11 & -.0073  & -.0114  &  .0095  &  .0115  \\
\hline
12 & -.0070  & -.0096  &  .0068  &  .0096   \\
\hline
13  & -.0044  & -.0072  &  .0078  &  .0079  \\
\hline
14 & -.0018  & -.0031  &  .0023  &  .0032  \\
\hline
15 & -.0029 & -.0053 &   .0060 $\ast$  & .0054  \\
\hline
16  & -.0022  & -.0026  &  .0026 $\ast$  &   .0030  \\
\hline
17  & -.0033  & -.0041  &  .0024  &  .0046 \\
\hline
18  & -.0081  & -.0100  &  .0079  &  .0094   \\
\hline
19 & -.0016  & -.0045  &  .0035  &  .0046  \\
\hline
20  & -.0054  & -.0080  &  .0079  &   .0080  \\
\hline
\end{tabular}
\end{center}
\small{ $\Delta = SVV_{PiCCO} - SVV_{SCSA}$  with $SVV_{PiCCO} = SV_{PiCCO}(n) - SV_{PiCCO}(n-1)$ and  $SVV_{SCSA} = SV_{SCSA}(n) - SV_{SCSA}(n-1)$.
$Min(\Delta)$  and $max(\Delta)$  stand for the minimal and maximal values of $\Delta$ respectively. $CI$ is 95\% confidence interval of the agreement limits of $\Delta$; $[CI_{inf} \Delta \:\: CI_{sup} \Delta]$. $\star$ means that all the values are inside $[CI_{inf} \delta \:\: CI_{sup}\Delta]$ except 1 value greater than $CI_{sup}\Delta$ for two subjects.}
\end{table}

\subsection{Cross-correlation analysis}
\begin{figure}[tb]
 \begin{center}
  \includegraphics[width=8cm]{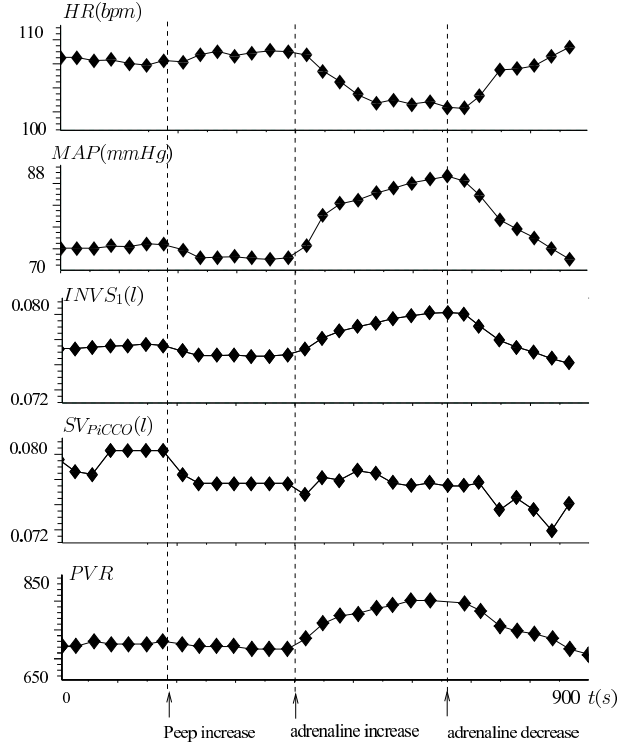}\\
  \caption{Cardiovascular time series of subject 21, submitted to ventilatory  and pharmacological changes. The  lack of correlation between $INVS_1$  and $SV_{PiCCO}$ (cross-correlation coefficient=0.039) is in favor of  $INVS_1$. Before increasing PEEP, nothing happens, $HR$, $MAP$, $PVR$  are stable, thus no $SV$ change can be expected. Nevertheless, $SV_{PiCCO}$  decreases then increases while $INVS_1$ remains quite stable.  During increasing adrenaline, increasing $PVR$ is accompanied, as expected,  by increasing $MAP$ and decreasing $HR$. An increase in $SV$ is also expected, which is done by $INVS_1$ while $SV_{PiCCO}$ remains quite stable.}\label{sujet21}
 \end{center}
\end{figure}

\begin{figure}[tb]
 \begin{center}
  \includegraphics[width=8cm]{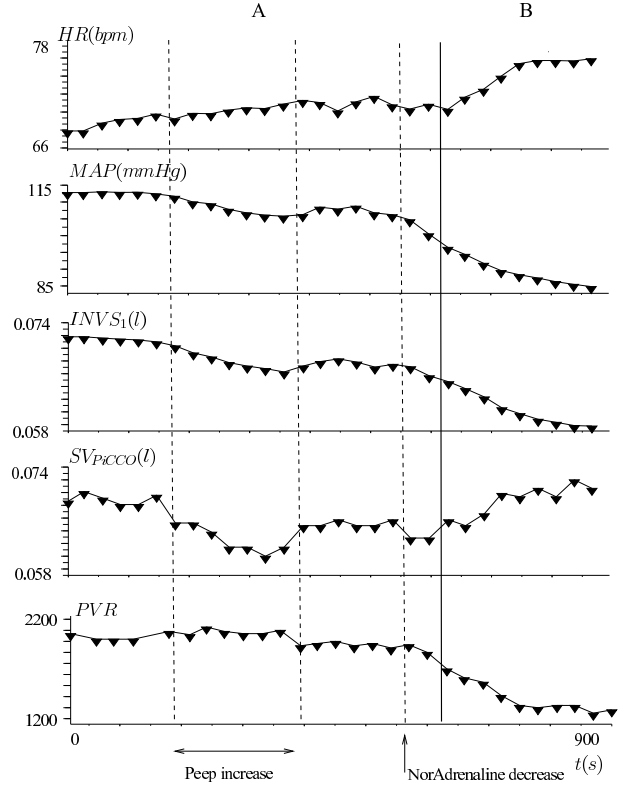}\\
  \caption{Cardiovascular time series for subject 17, submitted to ventilatory  and pharmacological changes. $INVS_1$ and $SV_{PiCCO}$  are strongly correlated (cross-correlation coefficient $= 0.84$) during period A, and they are divergent during period B. Decreasing noradrenaline is naturally accompanied by decreasing $PVR$ and $MAP$   and increasing $HR$. A decrease in $SV$ is also expected, which is  done by $INVS_1$ while $SV_{PiCCO}$ strongly increases.}\label{sujet17}
 \end{center}
\end{figure}

Table \ref{table1}, column three, shows the coefficients of cross-correlation in decreasing order. The amount of well correlated measures represents 94\% of the all recordings.
Because of an obvious divergence between $SV_{PiCCO}$ and $INVS_1$ (coefficients $< 0.1$),  Subject 21 and the third part of Subject 17 were discarded from the table. Thus, they represent only 6\% of discrepancy among all the recordings.
Fig.\ref{sujet21} and fig.\ref{sujet17} respectively represent the time series of $HR$, $MAP$, $PVR$, $INVS_1$ and $SV_{PiCCO}$ for  subjects 21 and 17.

\begin{figure}[tb]
 \begin{center}
  \includegraphics[width=8cm]{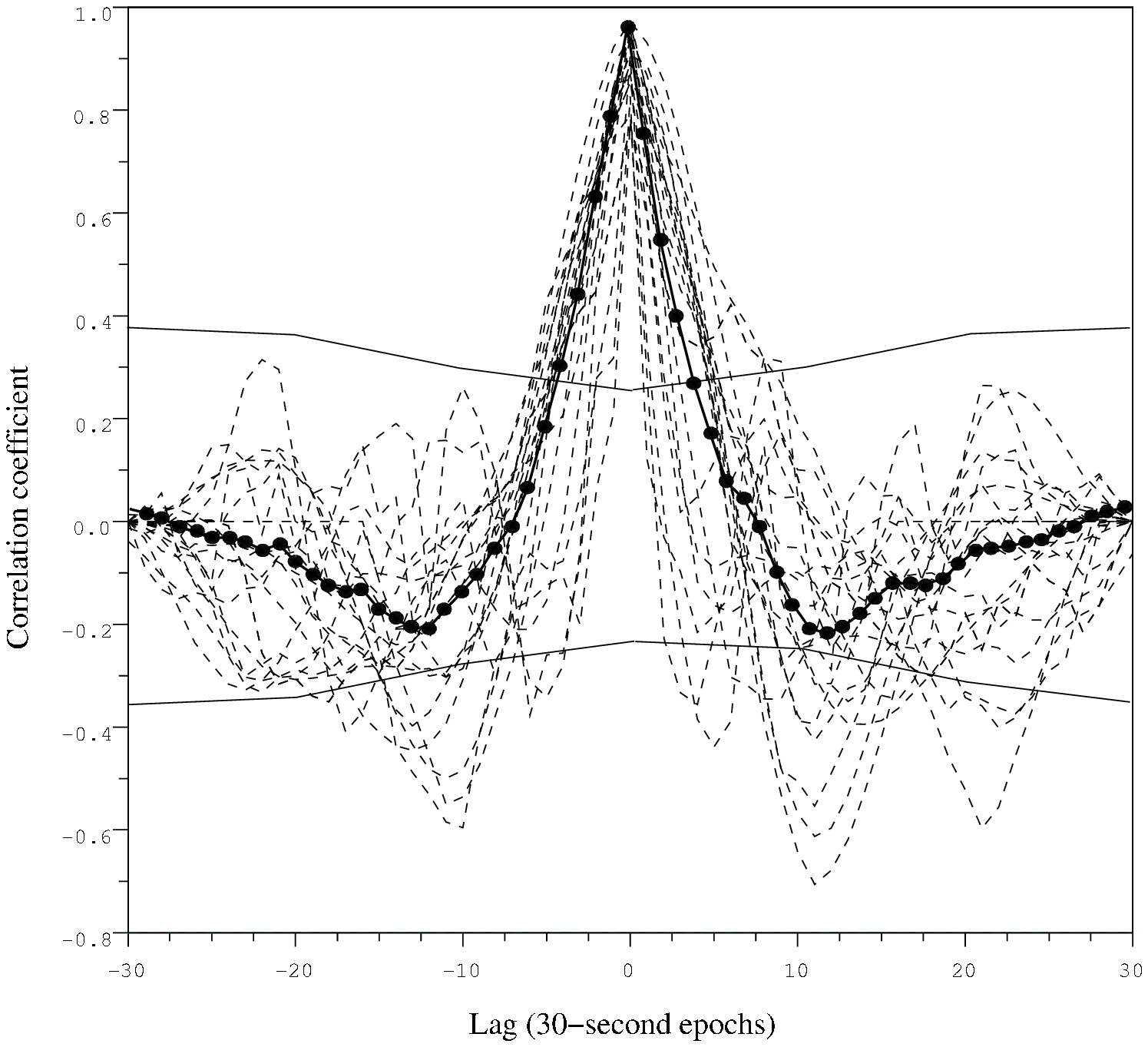}\\
  \caption{Cross-correlation between the 30-second averaged $SV_{PiCCO}$ and $INVS_1$ values in 20 subjects
Each line (dashed) stands for a subject; the strong line ($-\bullet-$) represents the average of the 20 subjects;
continuous lines represent critical $r$ values for a 5\% level of probability (Bravais-Pearson table). The vertical axis depicts the correlation coefficients. The horizontal axis depicts the lag, in number of 30-second measures, of one time series on another. The correlation coefficients of the unlagged data are shown at horizontal midpoint (lag 0). All time series are exactly synchronized, with a mean correlation coefficient equal to 0.9 at lag 0 (p=0.00001).
}\label{cross-correlation}
 \end{center}
\end{figure}

Fig.\ref{cross-correlation} represents the cross-correlation coefficients of the 20
remaining subjects (dashed lines). As homogeneity was verified, averaged values were also plotted ($- \bullet -$).
The vertical axis depicts the correlation coefficients. The horizontal axis depicts the lag, in number of 30-second
averaged values of one time series on another. The correlation coefficients of the unlagged data are plotted at horizontal midpoint (lag 0).
The two symmetrical continuous lines represent critical $r$ values for a level 5 of probability (Bravais-Pearson table). The greatest correlation stands at lag 0 for all the remaining subjects (mean correlation$= 0.90$; $sem=0.01$; $p=0.00001$). This result shows an excellent temporal similarity between the successive measures, indicating that they change in the same way over time.

\subsection{Linear regression}

Table \ref{table1}, column four, shows the $R$ coefficients of linear regression.  Cross-correlation and $R$ coefficients are strongly correlated (0.95 at Spearman rank order correlation test). The mean coefficient, equal to 0.89 shows a great degree of linearity between the two methods. Fig.\ref{figX} shows the plots of linear regression for the first three subjects.

\subsection{Bland-Altman method}

\begin{figure*}[htbp]
 \begin{center}
  \includegraphics[width=13cm]{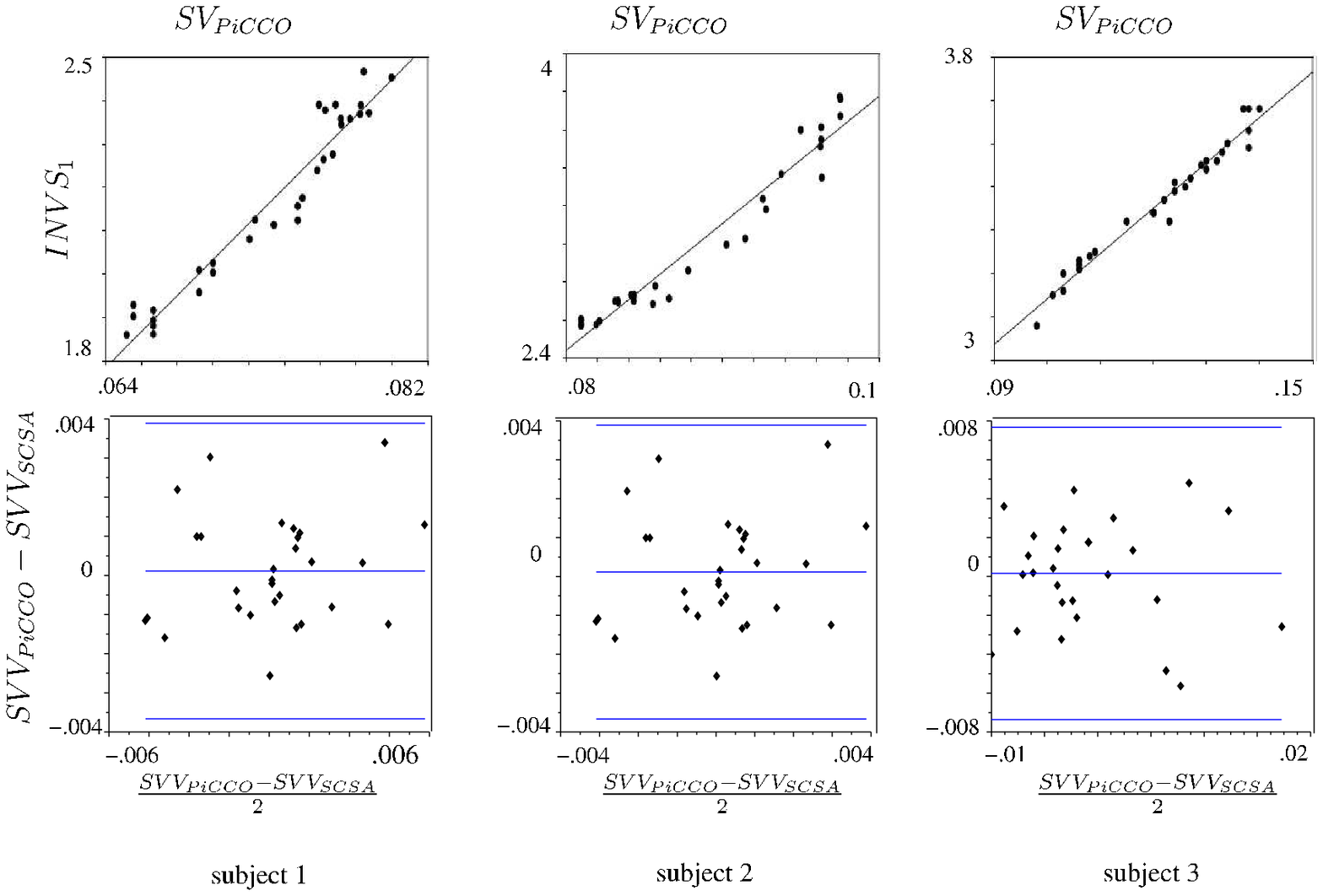}\\
  \caption{Linear regression plots (at the top) and Bland-Altman plots (at the bottom) for the three first subjects. The  coefficient of  correlation  is greater than  .95 for each of them, meaning a strong linear relation. All the differences in $SVV_{PiCCO}$ and $SVV_{SCSA}$ are included in the 95 \% confidence interval for each of them, meaning a good agreement between the two methods.
}\label{figX}
 \end{center}
\end{figure*}

Fig.\ref{figX} shows the Bland-Altman plots for the first three subjects. Differences in the two mean $SVV$ ($SVV_{PiCCO} - SVV_{SCSA}$) are
plotted against the mean of the two $SVV$  ($\dfrac{SVV_{PiCCO} + SVV_{SCSA}}{2}$).  Continuous lines respectively stand for the mean of
differences $\Delta$ between the two results, and the 95\% confidence interval. Each dot stands for the difference between $SVV$ measured by the two methods. In the three cases, mean $\Delta$ is equal to 0 and all values are within the two 95\%
confidence intervals. Table \ref{table2} gives results for all the subjects.  When comparing $min(\Delta)$ to $CI_{inf} \Delta$ (columns 1 and 2) and
$max(\Delta)$ to $CI_{sup}$ (columns 3 and 4),  one can see that all values are inside $[CI_{inf} \:\:\: CI_{sup}]$ except one value greater than
$CI_{sup} \Delta$ in two subjects ($\ast$). 
These global results allow us to conclude that a good proximity  exists between the two methods with the same order of dispersion and that they are  interchangeable.

\begin{figure}[tb]
 \begin{center}
  \includegraphics[width=8cm]{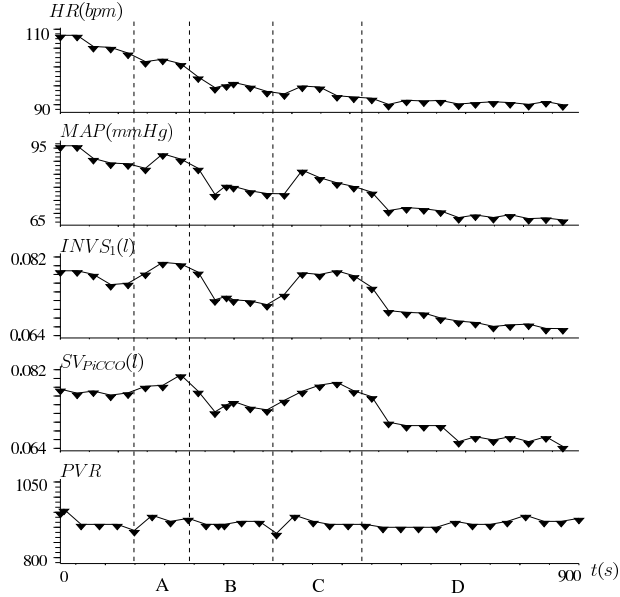}\\
  \caption{Cardiovascular time series of subject 1, submitted to Positive End Expiratory Pressure changes. A: decreasing PEEP; B: increasing PEEP; C: decreasing PEEP; D: increasing PEEP.  $INVS_1$ and $SV_{PiCCO}$  are strongly correlated (cross-correlation coefficient $= 0.99$).}\label{sujet1}
 \end{center}
\end{figure}

\begin{figure}[tb]
 \begin{center}
  \includegraphics[width=8cm]{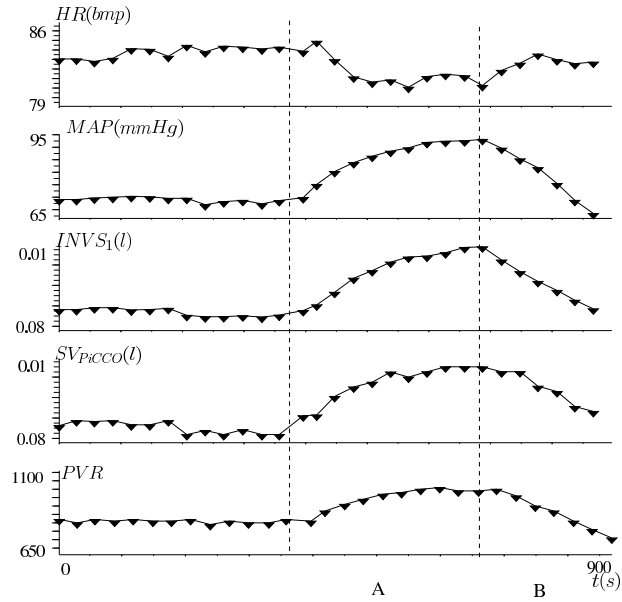}
  \caption{Cardiovascular  time series of subject 2, submitted to a vasoactive drug. A: increasing noradrenaline; B: decreasing noradrenaline.  $INVS_1$ and $SV_{PiCCO}$  are strongly correlated (cross-correlation coefficient $= 0.97$).}\label{sujet2}
 \end{center}
\end{figure}

\begin{figure}[tb]
 \begin{center}
  \includegraphics[width=8cm]{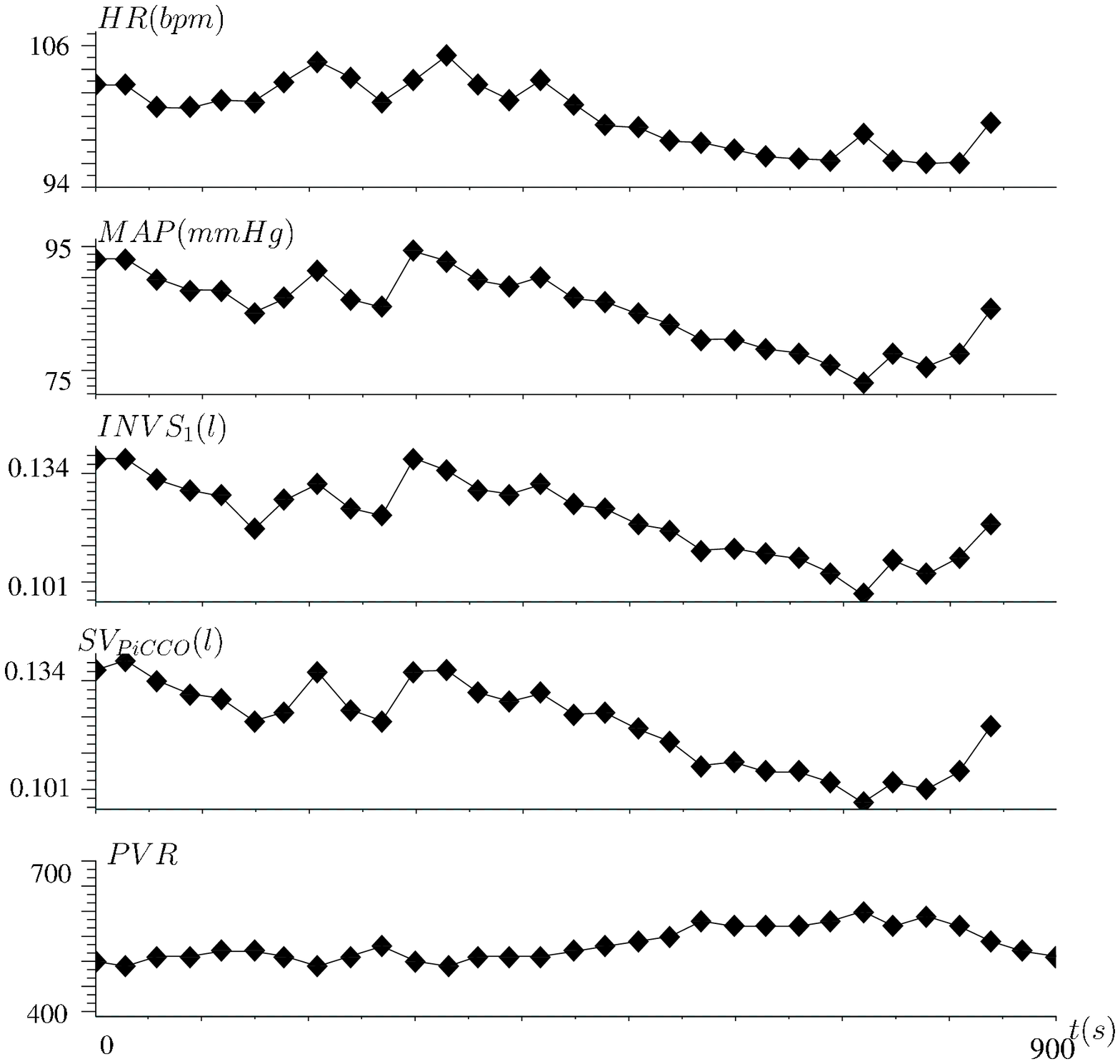}
  \caption{Cardiovascular time series of subject 3, without any change in ventilatory or pharmacological condition.  $INVS_1$ and $SV_{PiCCO}$  are strongly correlated (cross-correlation coefficient $= 0.97$).}\label{sujet3}
 \end{center}
\end{figure}

\section{Discussion}
A new method for a simple and minimally invasive  $SVV$ estimation from ABP measurements has been validated in this study. The ABP signal is reconstructed with a semi-classical signal analysis method SCSA which enables the decomposition of the signal into its systolic and diastolic parts.  Some spectral parameters, that give relevant physiological information,  are then computed and especially the first systolic invariant  $INVS_1$, given by the area under the estimated systolic pressure curve. Thus, we have shown that $INVS_1$ yields reliable $SVV$ assessment.

So, in order to validate this approach, we compared $INVS_1$ estimated from ABP measurements with $SV$ measured with a reference method: the PiCCO technique. Three statistical methods were applied for this validation: cross-correlation analysis, linear regression and the Bland-Altman test.
 Among the 315 minutes duration of all the 21 recordings, 94 \% presented a very high correlation between $INVS_1$ and $SV_{PiCCO}$. The mean coefficient was equal to 0.9 for cross-correlation, and equal to 0.89 for linear regression.  The remaining 6\% without correlation, concerned two subjects: all of subject 21 and the last third of subject 17. This discrepancy, interpreted with the help of the synchronized $HR$, $MAP$, $PVR$ time series, can be explained by the following  remarks:
\begin{itemize}

\item For subject 21 (fig.\ref{sujet21}),  before increasing PEEP, nothing happens, $HR$, $MAP$, $PVR$  are stable, thus no  $SV$ change can be expected. Nevertheless, $SV_{PiCCO}$  decreases then increases while $INVS_1$ remains quite stable.  During increasing adrenaline, increasing $PVR$ is accompanied, as expected,  by increasing $MAP$ and decreasing $HR$. An increase in $SV$ is also expected, which  is done by $INVS_1$ while $SV_{PiCCO}$ remains quite stable.
\item For the last third of subject 17 (fig.\ref{sujet17}, B), decreasing noradrenaline is naturally accompanied by decreasing $PVR$ and $MAP$   and increasing $HR$. A decrease in $SV$ is also expected, which is done by $INVS_1$ while $SV_{PiCCO}$ strongly increases.
\end{itemize}

The divergence between  $INVS_1$ and   $SV_{PiCCO}$ is in favor of  $INVS_1$  for these two subjects.

\begin{figure}[tb]
 \begin{center}
  \includegraphics[width=8cm]{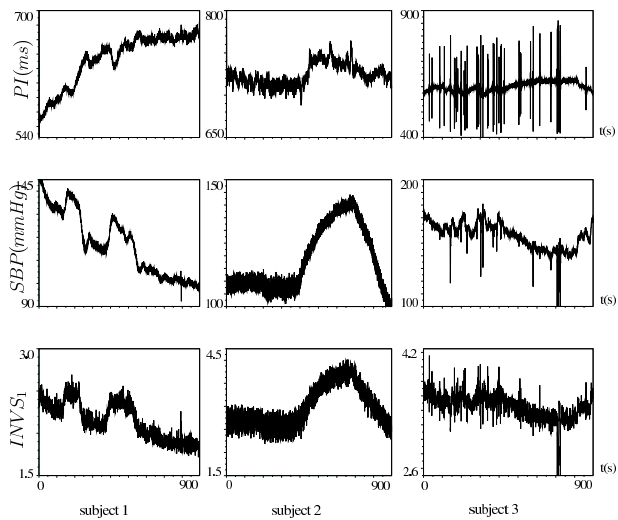}\\
  \caption{Pulse interval ($PI$), Systolic Blood Pressure ($SBP$) and $INVS_1$ time series for the three first subjects with a strong correlation between $SV_{PiCCO}$ and $INVS_1$. The  $INVS_1$ is precisely estimated  despite unstable hemodynamic conditions (subjects 2 and 3) and noisy conditions such as extra-systoles and artefacts (subject 3).
}\label{figY}
 \end{center}
\end{figure}

On the 94\% recordings with  well correlated  $INVS_1$ and   $SV_{PiCCO}$, the Bland-Altman test showed a very good agreement between the two approaches and demonstrated their interchangeability. It is worth noticing that this  agreement is  obtained  in  unstable hemodynamic and/or noisy conditions which prove the robustness of the SCSA method. Several great ventilatory or pharmacological changes are illustrated in  fig.\ref{sujet1}, fig.\ref{sujet2} and fig.\ref{sujet3} but also by the non averaged time series in fig.\ref{figY}. A  noisy condition is  illustrated by subject 3 on the right. Despite a raw $PI$ (top right) disturbed by extra-systoles and artefacts, $INVS_1$ (bottom right) is well estimated.

Therefore, this study shows that SCSA  is  a reliable method for $SVV$ assessment and more suitable in  the two cases of divergence. The good agreement between the two approaches could be explained by the fact that the main idea in the SCSA technique is quite similar to the PiCCO and consists in using the area under the systolic part of the pressure curve. However, the detection of the end systole with the SCSA is different from  the pulse contour approach. Indeed, while the pulse contour approach uses an algorithm to detect the dicrotic notch, the SCSA uses an ABP model based on solitons that takes into account nonlinear phenomena, as described in section II.C.1.b.  This difference could explain the greater reliability of SCSA when discrepancies between the two approaches appear. Moreover, this explanation agrees with the observation of the raw ABP signal  for subject 17: its shape is very different between  the first and last part of the recording.

Finally, unlike the PiCCO technique which needs periodic calibration by a thermodilution technique, SCSA is  easier to use, requiring  less equipments, only for  ABP measurements. It is much less invasive and could be totally noninvasive if we used a FINOMETER device, for instance. This latter point, already experimented  in our previous studies \cite{Laleg:08}, should be a new perspective for a simple non-invasive $SVV$ assessment.

\end{document}